\documentclass[preprint,showpacs,preprintnumbers,amsmath,amssymb,endfloats*]{revtex4}

\usepackage{graphicx}
\usepackage{dcolumn}
\usepackage{bm}

\begin{document}

\preprint{Casimir: Dissimilar materials}

\title{Measurement of the Casimir force between dissimilar metals}

\author{R. S. Decca$^{1}$, D. L\'opez$^{2}$, E. Fischbach$^{3}$, D. E.
Krause$^{4,3}$} \affiliation{$^{1}$Department of Physics, Indiana
University-
Purdue University Indianapolis, Indianapolis, IN 46202, USA\\
$^{2}$Bell Laboratories, Lucent Technologies, Murray Hill, NJ
07974, USA\\
$^{3}$Department of Physics, Purdue University, West
Lafayette, IN 47907, USA\\
$^{4}$Physics Department, Wabash College, Crawfordsville, IN
47933, USA}

\date{\today}

\begin{abstract}
The first precise measurement of the Casimir force between
dissimilar metals is reported. The attractive force, between a Cu
layer evaporated on a microelectromechanical torsional oscillator,
and an Au layer deposited on  an Al$_2$O$_3$ sphere, was  measured
dynamically with a noise level of 6~fN/$\sqrt{\rm{Hz}}$.
Measurements were performed for separations in the 0.2-2 $\mu$m
range. The results agree to better than 1\% in the 0.2-0.5 $\mu$m
range with a theoretical model that takes into account the finite
conductivity and roughness of the two metals. The observed
discrepancies, which are much larger than the experimental
precision, can be attributed to a lack of a complete
characterization of the optical properties of the specific samples
used in the experiment.
\end{abstract}


\maketitle

The Casimir force between two metallic layers arises from quantum
mechanical fluctuations of the vacuum  \cite{Casimir}.  In recent
years, an impressive amount of experimental
\cite{hobun,hobPRL,Mohideen} and theoretical
\cite{mostepanenko,theoretical} work has been performed to better
understand the Casimir force between real metals
 \cite{mostepanenko2,lamoreaux,lambrecht}.  While most of the work
has  focused on cases where the attracting bodies are composed of
the same material,  the theoretical models also include the case
of bodies composed of dissimilar metals  \cite{mostepanenko}.

Although the Casimir force is of fundamental importance in its own
right, it is also an unwanted background in current attempts to
search for new macroscopic forces over short distance scales
\cite{hypothetical}.  Such forces have been conjectured to arise
in unification theories, including those containing additional
spatial dimensions  \cite{ADD}.  It follows that setting stringent
limits on new forces from Casimir force measurements requires
either a very precise comparison between theory and experiment or
a method for suppressing the Casimir background  \cite{Ephraim}.

The preceding discussion provides the motivation for improving our
understanding of the Casimir effect.   More precise measurements
should result in better theoretical models which, in turn, will
yield a more complete picture of the Casimir effect, thus
improving our ability to detect new macroscopic forces.  Until
recently, experiments lagged behind theory;  with the development
of sensitive force transducers, however, it became necessary to
introduce refinements to the theory \cite{mostepanenko,lamoreaux}.

In this Letter, we report the first precise measurement of the
Casimir force between two dissimilar metals, at a precision $\sim
100$ times better than previous measurements.  At the current
noise level of $\sim 6$ fN/$\sqrt{\mbox{Hz}}$ our data shows a
small disagreement with the Lifshitz formula \cite{lifshitz},
which may be due to an incomplete characterization of the metallic
dielectric function, as suggested in \cite{lamoreaux}. Our results
thus suggest the need of additional experimental and theoretical
work to further improve their agreement for the Casimir effect.

A microelectromechanical torsional oscillator (MTO) has been used
since it is less affected by center of mass motions when compared
with other systems. A judicious selection of the geometry results
in a reduction of the spring constant $\kappa$ of the system by
over an order of magnitude. Furthermore, we used a dynamic scheme
which fully exploits the high quality factor $Q$ of the MTO.

The experimental setup is schematically shown in Fig. \ref{fig1}.
An Al$_2$O$_3$ sphere with a 600~$\mu$m nominal diameter was
sputter coated with a 1 nm layer of Cr followed by (203 $\pm$
6)~nm of Au. After coating, the sphere was glued with conductive
epoxy to the side of an Au coated optical fiber \cite{hobun}. The
radius of the coated sphere was measured to be (296 $\pm$ 2)
$\mu$m, the error arising because the Al$_2$O$_3$ ball was not
spherical. On the other hand, asymmetries induced by the
deposition process were measured to be smaller than 10 nm, the
resolution of the scanning electron microscope used. The
fiber-sphere assembly was moved vertically by a micrometer driven
stage in combination with a piezo-driven one.

The  MTO is made of a 3.5 $\mu$m thick, 500 $\times$
500~$\mu$m$^2$ heavily doped polysilicon plate suspended at two
opposite points by serpentine springs, as shown in the inset of
Fig.~\ref{fig2}. The springs are anchored to a SiN$_x$ covered Si
platform. Two independent polysilicon electrodes under the plate
allow the capacitance between the electrodes and the plate, which
are separated by  a gap of $2$ $\mu$m, to be measured.  We
calculated $\kappa = (wt^3E_{si}/6L_{serp}) \simeq 9.5 \times
10^{-10}$Nm/rad \cite{serpentine}, where $w = 2~\mu$m is the width
of the serpentine, $t = 2~\mu$m its thickness and $L_{serp} =
500~\mu$m its length. $E_{Si} = 180$ GPa is Si Young's modulus.
This value is in good agreement with the measured $\kappa = 8.6
\times 10^{-10}$~Nm/rad. The MTO is mounted on a piezo electric
driven $xyz$ stage which in turn is mounted on a micrometer
controlled $xy$ stage. The edges of the plate are coated with 1 nm
of Cr followed by 200 nm of Cu.

The combination of all translational stages allows positioning the
Au coated sphere on top of the metal coated platform. The
separation between the sphere and the platform is controlled by
the $z$-axis of the $xyz$ piezo stage. A two color fiber
interferometer based closed-loop system is used to keep the
fiber-to-platform separation $z_i$ constant. The separation
$z_{metal}$ between the two metallic surfaces is (see Fig.
\ref{fig1}): $ z_{metal} = z_i - z_o - z_g - b\theta$, where $b$
is the lever arm between the sphere and the axis of the MTO,
$\theta$ is the angle between the platform and the plate, and
$\theta \ll 1$ has been used. $z_o$ was measured
interferometrically by gently touching the platform with the
sphere. $z_g$ = (5.73 $\pm$ 0.08) $\mu$m, which includes the gap
between the platform and the plate, the thickness of the plate,
and the thickness of the Cu layer, was determined using an AFM.
The error in the interferometric measurements was found to be
$\Delta z_i^{rms} = 0.32$ nm, which incorporates the relative
vibrations of the fiber with respect to the platform and the
measurement noise. The entire system was kept at a pressure below
$10^{-4}$ torr.

The force between the two metallic layers is determined by
measuring the angle $\theta$ as a function of the separation
between the layers. Since $\theta \ll 1$, to a very good
approximation $\theta \propto \Delta C$, where $\Delta C$ is the
difference in capacitance between the two electrodes and the
platform. The proportionality constant is found by applying a
known potential difference between the two metallic layers at
separations larger than 3 $\mu$m. Thus, the net force can be
approximated by the electrostatic force $F_e$ \cite{smythe}

\begin{equation}
F_e = 2\pi\epsilon_o(V_{Au} - V_o)^2 \sum_{n=1}^{\infty}
\frac{[\coth (u) - n \coth (nu)]}{\sinh (nu)}, \label{eq1}
\end{equation}

\noindent where $\epsilon_o$ is the permittivity of free space,
$V_{Au}$ is the voltage applied to the sphere, $V_o$ the residual
potential difference between the metallic layers when they are
both grounded, and $\cosh u = [1+(z_{metal}+2\delta_o)/R]$. In
Eq.~(\ref{eq1}) it was found that only the first two terms of the
$(z_{metal}+2\delta_o)/R$ expansion give a significant
contribution. The average separation between the layers when the
metals come in contact, 2$\delta_o$,is determined by the roughness
of the films. As shown in Fig. \ref{fig2}, Eq.~(\ref{eq1}) can be
used to fit the measured force between the plane and the sphere as
a function of their separation for several values of $V_{Au} -
V_o$, where $V_o$ arises mainly from the difference between the
work functions of the two metals. Eq.~(\ref{eq1}) serves several
additional purposes:  i) It is used to calibrate the constant
between the measured $\Delta C$ and the applied force. ii) It is
used to find the voltage $V_{Au}$ where the electrostatic force is
zero, independent of the separation. iii) It is also used to fit
for the radius of the Au coated sphere $R = (294.3 \pm 0.1)$
$\mu$m, and iv) for the increase in the separation between the two
metallic layers, 2$\delta_o$. The fitted value of $\delta_o =
(39.4 \pm 0.3)$ nm agrees with the value estimated from AFM
measurements, $\delta_o = (33 \pm 9)$ nm. Furthermore, it was
found that the motion of the sphere parallel to the axis of the
MTO did not affect $\theta$, while its motion perpendicular to the
axis resulted in a linear dependence on $b$. We thus conclude that
there are no significant deviations from the assumption that the
Cu plane is of infinite extent.

The force between the two metals is $F = k \Delta C$, where
$\Delta C$ is measured to 1 part in 5 $\times$ 10$^5$ using a
bridge circuit  \cite{capacitance}. The constant $k$ is determined
by Eq.~(\ref{eq1}) to be $k = (50280 \pm 6)$~N/F. With this
approach the force sensitivity was found to be 1.4
pN/$\sqrt{\rm{Hz}}$. By setting $V_{Au} = V_o = 0.6325$ V, the
Casimir force between the Au coated sphere and the Cu planar film
was determined. The measurements are shown in Fig. \ref{fig3}(a).

To improve on  the force sensitivity, a direct use of the high
quality factor of the MTO was implemented  \cite{hobPRL}. In this
approach the vertical separation between the sphere and the
oscillator was changed as $\Delta z_{metal} =A \cos(\omega_r t)$,
where $\omega_r $ is the resonant angular frequency of the MTO,
and $A$ was adjusted between 3 and 35~nm for $z_{metal}$ equal to
0.2 and 1.2~$\mu$m, respectively. The solution for the oscillatory
motion yields \cite{hobPRL}
\begin{equation}
\omega_r = \omega_o \left [
1-\frac{b^2}{2I\omega_o^2}\frac{\partial F_c}{\partial z}\right ],
\label{approx}
\end{equation}
where $\omega_o \simeq \sqrt{\kappa/I}$ for $Q \gg 1$, and $I
\simeq 4.6 \times 10^{-17}$~kg~m$^2$ is the moment of inertia.
Since $A \ll z_{metal}$, terms of higher order in $\theta$
introduce a $\sim$ 0.1\% error. As before, Eq.~(\ref{eq1}) was
used to calibrate all constants. We found $\omega_o =
2\pi~(687.23$ Hz), and $b^2/2I = 6.489 \times 10^8$ kg$^{-1}$.
With an integration time of 10~s using a phase lock loop circuit
\cite{torsional}, changes in the resonant frequency of 10 mHz were
detectable. This allows the force to be calculated with a
sensitivity of the order of $\delta F_c$ = 6 fN/$\sqrt{\rm{Hz}}$,
a factor $\sim$ 3 larger than the thermodynamic noise. The results
for these measurements are plotted in Fig. \ref{fig4}.

In this configuration $\partial F_c/\partial z$ is being measured.
Using the proximity force theorem  \cite{Blocki}, the Casimir
force $F_{c}$ between a spherical surface of radius $R$ and an
infinite plane is given by $F_c = -\pi^3\hbar c R/360 z^3$
\cite{derivative}, where $c$ is the speed of light in vacuum and
$\hbar$ is Planck's constant. Its derivative $\partial
F_{c}/\partial z = 2\pi R P_{c}$, where $P_{c}$ is the force per
unit area between two infinite planes.

When finite conductivity effects are taken into account, the
Casimir force per unit area between two planes $P_{CP}$, and the
force between a sphere and a plane $F_{CS}$, are given by
\cite{lifshitz,mostepanenko}
\begin{subequations}
\label{eq2}
\begin{eqnarray}
P_{CP} &= -\frac{\hbar}{2\pi^2c^3}\int_0^\infty
\xi^3d\xi\int_1^\infty p^2 \left\{  \left [
\frac{(s_1+p)(s_2+p)}{(s_1-p)(s_2-p)} e^{2p\xi
z/c}-1 \right ]^{-1} \right.\nonumber \\
&+  \left. \left [\frac{(s_1+\epsilon_1p)(s_2+\epsilon_2p)}
{(s_1-\epsilon_1p)(s_2-\epsilon_2p)} e^{2p\xi z/c}-1 \right
]^{-1}\right\} dp, \label{eq2a}
\end{eqnarray}
\begin{eqnarray}
F_{CS} &= \frac{\hbar}{2\pi c^2}R\int_0^\infty \xi^2
d\xi\int_1^\infty p\left\{\ln\left
[1-\frac{(s_1-p)(s_2-p)}{(s_1+p)(s_2+p)}
e^{-2p\xi z/c}\right ]\right.\nonumber \\
&+\left.\ln \left [ 1-\frac{(s_1-\epsilon_1p)(s_2-\epsilon_2p)}
{(s_1+\epsilon_1p)(s_2+\epsilon_2p)} e^{-2p\xi z/c}\right
]\right\} dp, \label{eq2b}
\end{eqnarray}
\end{subequations}
where $\epsilon_j(i\xi)$ is the dielectric function of metal $j$,
$\omega = i\xi$ is  the complex frequency, and $s_j =
\sqrt{\epsilon_j-1+p^2}$. Since the metals are not smooth, the
expressions in Eq.~(\ref{eq2}) should be averaged over different
possible separation distances determined by the surface roughness
\cite{romero},
\begin{subequations}
\label{eq4}
\begin{equation}
P_C=\sum_i w_i P_{CP}(z_i), \label{eq4a}
\end{equation}
\begin{equation}
F_C=\sum_i w_i F_{CS}(z_i), \label{eq4b}
\end{equation}
\end{subequations}
to model the data shown in Figs.~\ref{fig3} and \ref{fig4}. The
probabilities $w_i$ were found using the AFM profiles of the
interacting surfaces. The values of the dielectric constant for Au
and Cu were obtained from \cite{palik}, extended to low energies
using a Drude model \cite{lambrecht}. The agreement between model
and data is better than 1\% for all cases where $\delta F(z) \leq
0.01 F(z)$, {\em i.e.} $z < 0.5~\mu$m.

There are, however, measurable differences between the
experimental and theoretical values of the Casimir energy.
Although the differences are larger than the experimental error,
they likely  arise from the $\epsilon(\omega)$ used, not
necessarily reflecting shortcomings of Eq. (\ref{eq2}): Using
different optical constants for Au and Cu \cite{Auconst}, or
different models of roughness, results obtained from
Eq.~(\ref{eq4}) vary by more than 1\%.

Other possible sources of discrepancy are:  i) finite thickness of
the metallic layers, and ii) effects of temperature. Since the
layers are much thicker than the penetration depth of
electromagnetic radiation on the system \cite{mostepanenko}, we
can treat them as effectively infinite. Finite temperature
corrections have been found to be $\lesssim 0.1\%$ using a plasma
dispersion model \cite{mostepanenko}. Other approaches
\cite{TEmode} predict corrections of up to 15\%. Although these
corrections have the same sign as the deviation observed in our
experiment, our data do not support their results. The sensitivity
is such that the effect of thermal corrections should be taken
into account once the metals are fully characterized.

Given the improved sensitivity of our measurement of the Casimir
force, we can set more stringent limits on new macroscopic forces
acting over short separations. A new force would appear as a
perturbation to the usual gravitational potential energy
\cite{Ep-book}. Such perturbation, a Yukawa interaction, is
characterized by its strength $\alpha$ and its range $\lambda$.
Using the differences of Fig. \ref{fig3}b as a benchmark for the
manifestation of new forces, we can set a limit $\alpha(\lambda =
200~\rm{nm}) \sim 10^{13}$, which represents an improvement of a
factor of 4 with respect to previously reported limits
\cite{Ephraim}. If, once the metals are properly characterized,
the difference between the measured and calculated values of the
Casimir force is found to be on the order of the experimental
error, then at the 1~$\sigma$ level our results give
$\alpha(200~\rm{nm}) \leq 10^{11}$. These limits can be improved
by directly comparing the force between the sphere and two
different metals, or two different isotopes of the same element
\cite{us}.  Such measurements are presently underway, and will be
presented elsewhere.

In summary, it was shown that by using a MTO, the sensitivity in
the force between two metallic films is improved by $\sim$~100
over previous experiments. This highlights the necessity of a
simultaneous measurement of the dielectric constant of the
metallic films used in the force measurements, and a better
understanding of the Casimir force between non-ideal bodies. When
used to set new limits in the search for new forces, it is
concluded that an improvement of more than 2 orders of magnitude
can be achieved for separations $\sim$ 100 nm. The dynamic method
employed also allows for a direct experimental test of the
proximity force theorem.

We are deeply indebted to G. L. Klimchitskaya and V. M.
Mostepanenko for help with the theoretical analysis. We also wish
to thank H. B. Chan, A. Lambrecht, S. Reynaud, and U. Mohideen for
useful discussions. RSD acknowledges financial support from the
Petroleum Research Foundation through ACS-PRF\#37542-G. The work
of EF is supported in part by the U.S. Department of Energy under
Contract No. DE-AC02-76ER071428.

\newpage

\begin{figure}
\caption{\label{fig1} Schematic of the experimental setup showing
its main components. The inset shows an schematic of the
electronic circuit used for the static measurements.}
\end{figure}

\begin{figure}
\caption{\label{fig2} Electrostatic force as a function of
separation for $V_{Au} - V_o = 0.24$ V and 0.30 V. Inset: Scanning
electron microscope image of the MTO showing the serpentine
spring.}
\end{figure}

\begin{figure}
\caption{\label{fig3} Casimir force as a function of separation.
The separation between the metallic layers has been adjusted to
account for the roughness: $z = z_{metal}+2\delta_o$ {\bf(a)}
Direct measurement of the force. The solid line is a fit using
Eq.~(\ref{eq4b}). {\bf(b)} Experimental data subtracted from the
theoretical model.}
\end{figure}

\begin{figure}
\caption{\label{fig4}{\bf(a)} Derivative of the Casimir force (see
text) as a function of separation. The solid line is a fit using
Eq.~(\ref{eq4a}). {\bf(b)} Experimental data subtracted from the
theoretical model. The deviation with respect to the model at
small separations is partially associated with non-linear terms on
Eq.~(\ref{approx}).}
\end{figure}

\end{document}